\documentclass[preprint,12pt]{elsarticle}

\usepackage{graphicx}
\usepackage{epsfig}

\usepackage{amssymb}

\journal{Physica C Special Volume on M2S IX 2009 (Tokyo) proceedings}

\begin{document}

\begin{frontmatter}



\title{Observation of Vortex Matching Phenomena in Antidot Arrays of NbN Thin Films}


\author[nims]{A. D. Thakur\corref{cor1}\fnref{fn1}}
\ead{thakur.ajay@nims.go.jp;adthakur@gmail.com}
\author[nims]{S. Ooi\fnref{fn1}}
\author[tifr]{S. P. Chockalingham}
\author[tifr]{John Jesudasan}
\author[tifr]{P. Raychaudhuri}
\author[nims]{K. Hirata}
\cortext[cor1]{Corresponding author}
\fntext[fn1]{ADT and SO would like to acknowledge partial support from World Premier International Research Center (WPI) Initiative on Materials Nanoarchitectonics, MEXT, Japan.}
\address[nims]{National Institute for Materials Science, 1-2-1 Sengen, Tsukuba, Ibaraki 305-0051, Japan}
\address[tifr]{Tata Institute of Fundamental Research, Homi Bhabha Road, Colaba, Mumbai 400005, India}



\begin{abstract}
We report vortex matching phenomenon in rectangular antidot array fabricated on epitaxial NbN thin film. The antidot array was fabricated using Focussed Ion Beam milling technique. The magneto-transport measurements points to a period doubling transition at higher magnetic field for rectangular lattices. The results are discussed within the light of several models including the multi-vortex model, the matched lattice model and the super-matched lattice model.

\end{abstract}


\end{frontmatter}

\section{Introduction}
\label{sec1}
Among the low transition temperature ($T_{\rm c}$) superconductors, NbN has a relatively high $T_{\rm c}$ ($\sim 16~K$). It has a small coherence length ($\xi_0 \sim 5~nm$), a large penetration depth ($\lambda \sim 200~nm$), is chemically inert to ambient atmospheric environment, and is amenable to various sub-micron fabrication techniques. Thus NbN presents itself as a suitable system for exploring vortex physics phenomena both from academic as well as applied perspective. Interesting phase-slip experiments have been reported on mesoscopic superconducting NbN wires \cite{Peet}. However, owing to typical values of its mean free path ($l$) and Fermi wave vector ($k_F$), it falls in the limit of dirty superconductors in accordance with the Ioffe-Regel criteria \cite{I1}. Therefore NbN has been considered to have a high degree of bulk pinning, and, until recently \cite{apl2009b}, it has not been possible to observe vortex matching phenomena in samples of NbN with artificial pinning centers. Vortex matching phenomena have been reported in a large variety of low $T_{\rm c}$ \cite{fiory, baert, silva,tatbat},  and high $T_{\rm c}$ superconductors \cite{ooi} in the past. Recently we demonstrated the occurrence of robust vortex matching phenomena in engineered thin films of NbN \cite{apl2009b}. In this paper we extend our work to rectangular antidot arrays.
\section{Experimental}
\label{expt}
The rectangular antidot array sample was fabricated on NbN thin films grown on a (100) MgO substrate. A 60 nm thick NbN film was deposited through reactive dc sputtering using a Nb target in (1:5) Ar-N$_2$ partial pressure of 5 mTorr keeping the substrate at 600$^o$C \cite{P1}. The film had a superconducting transition ($T_{\rm c}^{on}$) of 15.01 K, a coherence length ($\xi_0$) of 4.3~nm and a $k_Fl$ of 5.9 \cite{P1}. Using a mask-aligner, a probe pattern with 40$\times$40~$\mu$m$^2$ wide regions between the voltage probes is then made via photolithography followed by ion beam etching. Rectangular antidot arrays were fabricated within these 40$\times$40~$\mu$m$^2$ wide regions using the Focussed Ion Beam (FIB) milling technique utilizing an aperture of 25~$\mu$m and a typical ion-beam current of 4.1 pA using the {\it Micrion 2100 FIB}.   A typical $T_{\rm c}$ reduction of about 100~mK was observed after the fabrication of the antidot arrays. The temperature stability was better than 1~mK during the magneto-transport measurements.
\section{Results and Discussion}
\label{randd}
The fabricated rectangular antidot array had lattice spacings of 300 nm and 600 nm for the shorter and the longer axes respectively and the average antidot diameter ($D$) was $\sim$170~nm. The dc current ($I_{\rm dc}$) for the electrical transport measurements was applied along the longer axis of the array. The magnetic field was applied perpendicular to the plane of the film. Figure 1 shows the plot of resistance ($R$) versus filling fraction ($f=\phi / \phi_0=H/H_1$) at $I_{\rm dc}$ of 150~$\mu$A obtained at a temperature of 13.85 K. Here, $H$ ($\phi$) is the applied field (magnetic flux) and $H_1$ is the first matching period determined by a situation where there is a single quantum of flux $\phi_0$ corresponding to each rectangular plaquette (unit cell of the antidot lattice). In the present case $H_1=$ 115~$Oe$. As can be observed, $R$ has a notable modulation with distinct local minima at various values of $f$ ($=$ 1, 2, 3, 5, 7, 9, 11). These local minima correspond to the various matching periods and at high fields are less well defined with their periodicity apparently guided by the antidot array pitch along the Lorentz force direction \cite{S1}. One can also observe the signatures of fractional matching at $f= \frac{1}{2}, \frac{3}{2},~and~\frac{5}{2}$. However the fractional matching periods are absent at the higher field end. The inset of Fig. 1 shows the scanning ion image (SIM) of a portion of the rectangular antidot array.

Within the scenario proposed by Mkrtchyan and Shmidt (M$\&$S) \cite{MS1}, the maximum number of vortices that can be captured by an isolated antidot (or, a columnar defect) with a diameter $D$ is given by the saturation number ($n_{\rm s}$) such that, $n_{\rm s} = D/4 \xi (t)$, where $\xi (t)$ is the coherence length at reduced temperature $t (=T/T_{\rm c})$ given by $\xi (t) = \xi_0/ \sqrt{(1-t)}$, with $\xi_0$ being the zero temperature coherence length \cite{MS1}. This means that a maximum of $n_{\rm s}$ vortices can be captured by the antidots (multivortex scenario). In the case of the rectangular antidot array with $d=170~nm$, $n_{\rm s} \approx 2.6$ at 13.85 K considering a value of 4.3 nm for $\xi_0$ \cite{P1}. Doria {\it et. al.} \cite{doria} took into consideration the vortex-vortex interaction and suggested that $n_{\rm s} \sim (D/2\xi(t))^2$. This leads to a relatively larger estimate for $n_{\rm s}$ ($\sim$ 27) in the present case. However, it should be noted that a microscopic estimate of the proportionality constant ($C$) in Doria {\it et al.,} \cite{doria} estimate ($n_{\rm s} = C\cdot(D/2\xi(t))^2$) is absent and it is expected to depend on both the lattice parameters as well as the lattice symmetry of the antidot array. In the case of the current experiment, within the multi-vortex (MV) scenario, all vortices up to $f=2$ are captured at the antidots and the 3$^{rd}$ vortex nucleates at the interstitial site in the middle of the unit cell in the case of $f=3$. For $f>3$, there appears to be a crossover/transition to a scenario where vortices nucleate at the interstitial sites ramifying into a situation where the smaller axes of the antidot array appears to guide the values of the further matching periods. This leads to a period doubling transition at higher fields in the present case, where, $b/a=2$, with $a$ and $b$ being the unit cell lengths perpendicular and parallel to the current flow direction, respectively. The vortex-vortex interaction \cite{doria} is expected to affect the estimation of $n_{\rm s}$ within the M$\&$S scenario. Thus, experimentally observed $n_{\rm s}$ could be larger than the M$\&$S estimate. To the best of our knowledge, there does not exist a clear theoretical understanding of the phenomena. Bezryadin and Pannetier observed such a co-existence between multi-quanta vortices and interstitial vortices in Bitter decoration experiments on antidot arrays \cite{BP}. On the other hand, within a simple matched lattice (ML) model \cite{mart1} each antidot site can only accommodate a single vortex and the other vortices nucleate at the interstitial locations. Within ML model, the vortex lattice reconfigures from a commensurate rectangular to a square configuration  as soon as the vortex elastic energy ($E_{elastic}$) dominates over the pinning energy ($E_{pinning}$) \cite{mart2}. But, it predicts contrary to the observations in the present case, a quadratic (with respect to the number of the matching period) dependence instead of a the linearly equally spaced (in field) of the matching periods at the high field end \cite{S1}. Matched lattice model with an additional reorientation mechanism, also known as the super-matched lattice (SL) model seems to be in agreement with the observed phenomena \cite{Met1,RH1}. Our earlier studies on square and triangular antidot arrays \cite{apl2009b} seem to be in good agreement with the numerical simulations within the SL model done by Reichhardt {\it et al.,} \cite{RH1}. However, no such simulations exist for the rectangular antidot array case. A detailed local imaging study e.g., magneto-optical imaging (MOI) is required to distinguish between the validity of the MV and the SL scenarios in the present case. 

\section{Summary}
We observe the vortex matching effect phenomena in rectangular antidot array sample fabricated on NbN thin film.  There is a period doubling as one goes from the low field to the high field region. The fractional matching periods are absent at the high field end. The experimental data suggest an agreement with the matched lattice model with an additional reorientation mechanism \cite{Met1}.

\newpage

\begin{figure}
\includegraphics[scale=0.5,angle=0]{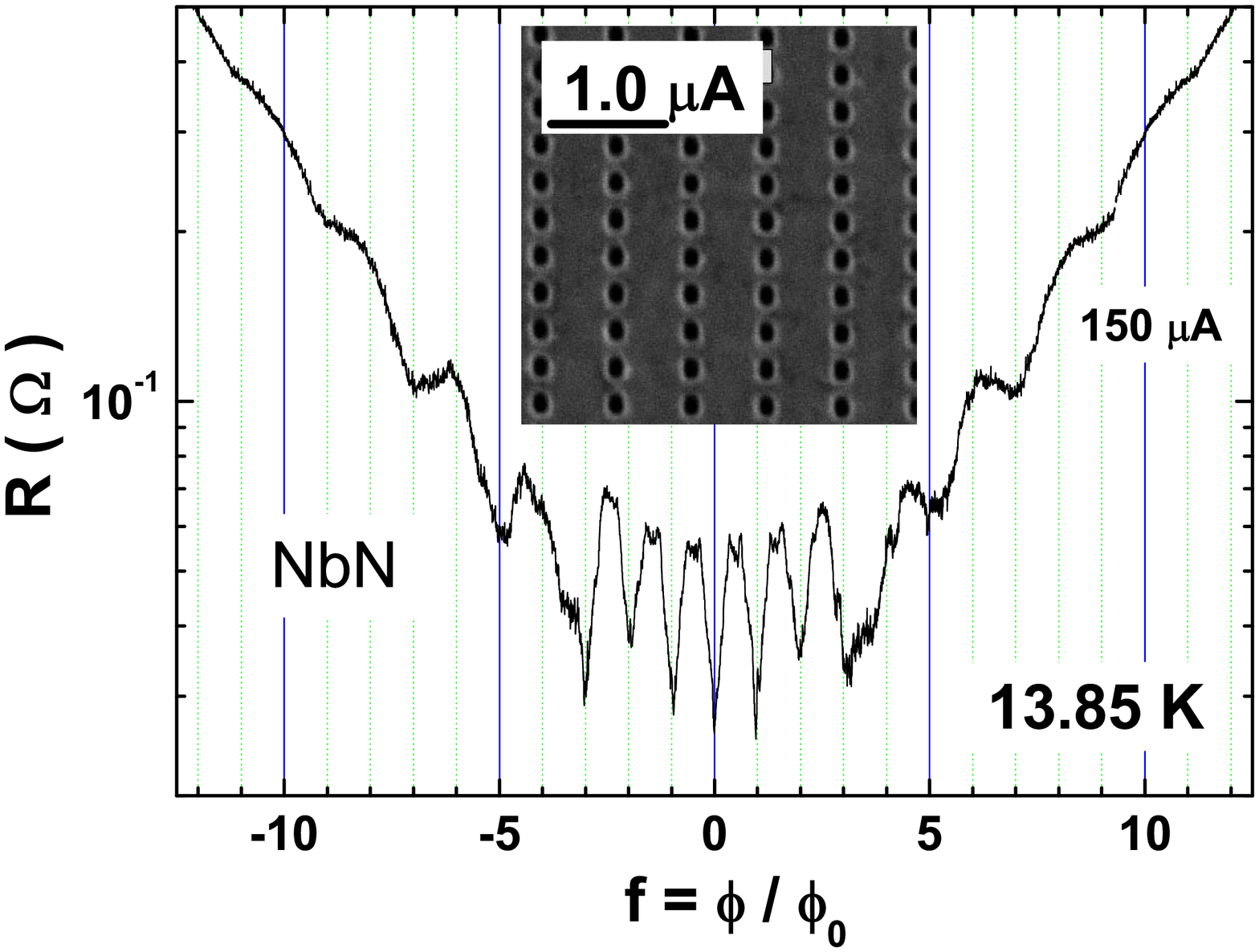}
\caption{{\it R} versus {\it f} data obtained at a drive current of  $I_{\rm dc}=150~\mu$A for a rectangular antidot lattice observed at a temperature of 13.85 K. The inset panel shows the SIM image of a portion of the rectangular antidot lattice with a pitch of 300~nm$\times$600~nm and an average antidot diameter of 170~nm.}
\end{figure}

\end{document}